\documentclass[letterpaper, twocolumn,superscriptaddress, showpacs,preprintnumbers,amsmath,amssymb]{revtex4}

\usepackage{graphicx}
\usepackage{dcolumn}
\usepackage{bm}

\begin{document}

\title{The many-body localization phase transition} 

\author{Arijeet Pal}
\affiliation{Physics Department, Princeton University, Princeton, NJ 08544, USA}

\author{David A. Huse}
\affiliation{Physics Department, Princeton University, Princeton, NJ 08544, USA}

\begin{abstract}
We use exact diagonalization to explore the many-body localization transition in a random-field spin-1/2 chain.
We examine the correlations within each many-body eigenstate,
looking at all states and thus effectively working at infinite temperature.  For weak random field the
eigenstates are thermal, as expected in this nonlocalized, ergodic phase.  For strong random field the eigenstates are localized,
with only short-range entanglement.  We roughly locate the localization transition and examine the probability distributions of the correlations,
finding that this quantum phase transition at nonzero temperature may be 
showing infinite-randomness scaling.
\end{abstract}


\maketitle


As originally proposed in Anderson's seminal paper \cite{pwa58}, 
an isolated quantum system of many interacting degrees of freedom with quenched disorder may be localized,
and thus generically fail to approach local thermal equilibrium, even in the limits of long time and large systems.
In the same paper, Anderson also treated the localization of a single particle-like quantum degree of freedom,
and it is this single-particle localization, without interactions, 
that has received the vast majority of the attention in the half-century since then.
Much more recently, Basko, {\it et al.} \cite{baa} have presented a very thorough study of many-body localization with interactions, 
and the topic is now beginning to receive more attention; see e.g. \cite{OHED,zpp,oph,mm,im,aas,mg}.
The many-body localization phase transition is of very fundamental interest,
since it is a quantum ``glass transition'' where equilibrium quantum statistical mechanics breaks down:
in the localized phase the system fails to thermally equilibrate.
This transition may be underlying some highly nonlinear low-temperature current-voltage characteristics measured in certain thin films \cite{akla}.
Also, it may soon be produced and studied in systems of ultracold atoms in random optical potentials \cite{bdz}.

Many-body localization appears to occur for a wide variety of particle and spin models.
Anderson's original proposal was for a spin system \cite{pwa58}, and the specific model we look at here is also a spin model,
namely the Heisenberg spin-1/2 chain with random fields along the $z$-direction \cite{zpp}:
\begin{equation}
H = \sum_{i=1}^L [h_i \hat{S}_i^{z} + J\hat{\vec{S_i}} \cdot \hat{\vec{S}}_{i+1}]~,
\end{equation}
where the static random fields $h_i$ are independent random variables, with a probability distribution that is uniform in $[-h, h]$.  
Except when we state otherwise, we will take $J=1$.  We consider chains of length $L$, with periodic boundary conditions.
This is one of the simplest models that shows a many-body localization transition.  Since we will be studying the system's behavior by exact diagonalization,
working with this one-dimensional model that has only two states per site 
allows us to probe longer length scales than would be possible for models on higher-dimensional lattices or with more states per site.
Fortunately, in the limit $L\rightarrow\infty$ the many-body localization transition at $h=h_c$ does appear to occur in this one-dimensional model.
The usual arguments that forbid phase transitions at nonzero temperature in one dimension do not apply here,
since those arguments rely on the validity of equilibrium statistical mechanics, which is exactly what is failing at the localization transition.

Our model has two global conservation laws: total energy, which is conserved for any isolated quantum system with a time-independent Hamiltonian;
and total $\hat{S^z}$. The latter conservation law is not essential for localization,
although its presence might possibly affect the universality class of the phase transition.
For convenience, we restrict our attention to states with zero total $\hat{S^z}$.
This restriction does enter in determining finite-size effects, but we see no reason for it to affect the nature of the phase transition.

For simplicity, we focus on the case of infinite temperature, where all states are equally probable
(and where the sign of the spin-spin interaction $J$ does not matter).
The localization transition also occurs at non-infinite temperature;
by working at infinite temperature we remove one parameter from the problem and are able to use all the data obtained from
a full exact diagonalization (within the zero total $\hat{S^z}$ sector) of each realization of our model Hamiltonian.
We see no reason to suspect that the nature of the localization transition is different between infinite and finite nonzero temperature,
although the transition at strictly zero temperature \cite{gs} is certainly different.
It is very important to emphasize that what we are studying is a quantum phase transition that occurs at nonzero (even infinite) temperature.
Like the more conventional ground-state quantum phase transitions, this transition is a sharp change in the properties of the many-body eigenstates of the Hamiltonian,
as we discuss below.  But unlike ground-state quantum phase transitions,
the many-body localization transition at nonzero temperature appears to be only a dynamical phase transition
and to be invisible in the equilibrium thermodynamics \cite{OHED}.

There are many distinctions between the many-body localized phase at large random field $h>h_c$
and the delocalized phase at small random field $h<h_c$.
We will call the latter the ``ergodic'' phase, although precisely how ergodic it is remains to be determined.
The distinctions between the two phases all have at their root the properties of the many-body eigenstates of the Hamiltonian,
which of course enter in determining the dynamics of the isolated system.

In the ergodic phase, the many-body eigenstates are thermal \cite{deutsch,sred,tasaki,rdo},
which permits the isolated quantum system to relax to thermal equilibrium under the dynamics due to its Hamiltonian.
In the thermodynamic limit ($L\rightarrow\infty$), {\it the system thus successfully serves as its own heat bath} in the ergodic phase.
In a thermal eigenstate, the reduced density operator of a finite subsystem converges to the equilibrium thermal distribution for $L\rightarrow\infty$.
And thus in a many-body eigenstate in the ergodic phase, the entanglement between a finite subsystem and the remainder of the system is,
for $L\rightarrow\infty$, the thermal equilibrium entropy of the subsystem.
At nonzero temperature, this entanglement is extensive, proportional to the number of degrees of freedom in the subsystem.

In the many-body localized phase at $h>h_c$, on the other hand, the many-body eigenstates are not thermal \cite{baa}:
the ``Eigenstate Thermalization Hypothesis'' \cite{deutsch,sred,tasaki,rdo} is false in the localized phase.
This means that in the localized phase, the isolated quantum system {\it does not relax to thermal equilibrium} under the dynamics of its Hamiltonian.
The infinite system fails to be a heat bath that can equilibrate itself.  It is a ``glass'' whose local configurations are set by the initial conditions.
Here the eigenstates do not have extensive entanglement, which makes them accessible to DMRG-like numerical techniques \cite{zpp}.
A limiting case of the localized phase that is simple is $J=0$ with $h>0$.  Here the spins do not interact,
all that happens dynamically is local Larmor precession of the spins about their local random fields.  No transport
of energy or spin happens, and the many-body eigenstates are simply product states with each spin either ``up'' or ``down''.

Any nonequilibrium initial condition is a mixed state that can be written as a density matrix in terms of the
Hamiltonian's eigenstates as: $\rho=\sum_{mn}\rho_{mn}|m\rangle\langle n|$.  The eigenstates have different energies,
so as time progresses the off-diagonal density matrix elements $m\neq n$ dephase from the particular phase
relations that determined the initial conditions, while the diagonal elements $\rho_{nn}$ do not change with time.
In the ergodic phase for $L\rightarrow\infty$, where all the eigenstates are thermal, this
dephasing necessarily brings any finite subsystem to thermal equilibrium.  But in the localized phase the eigenstates are all locally
different and athermal, so local information about the initial condition is also stored in the diagonal density
matrix elements, and it is the permanence of this information that in general prevents the isolated quantum system from relaxing to
thermal equilibrium in the localized phase.

Our goals in this paper are firstly to present exact diagonalization results in the ergodic and localized phases,
showing that they are consistent with the expectations discussed above.  Secondly, and more importantly, we examine some of the
behavior of the many-body eigenstates of our finite-size systems in the vicinity of the localization transition to see
what we can learn about the nature of this phase transition.  Although the many-body localization transition has been
discussed by a few authors, there does not appear to yet be any proposals for the nature (the universality class) of this phase
transition or for its finite-size scaling properties, other than some very recent initial proposals in Ref. \cite{mg}.
It is our purpose here to investigate these questions,
extending the previous work of Oganesyan and Huse \cite{OHED}, who looked at the
many-body energy-level statistics of a related one-dimensional model.
Since the many-body eigenstates have extensive entanglement on the ergodic side of the transition, it may be that
exact diagonalization (or methods of similar computational ``cost'' \cite{mg})
is the only numerical method that will be able to access the properties of the eigenstates on that side
of the transition.







\begin{figure}[!hbtp]
\includegraphics[height=2.2in,width=3.0in]{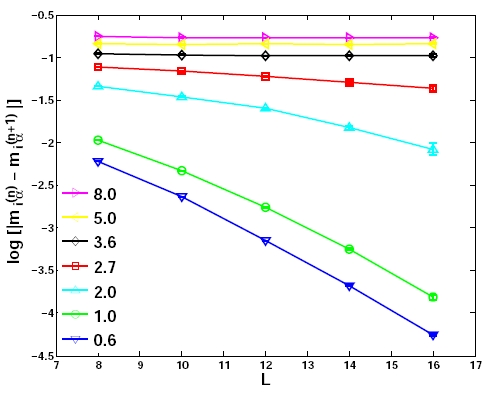}
\caption{(Color online) The logarithm of the mean difference between the local magnetizations in
adjacent eigenstates (see text).  The values of the random field $h$ are indicated in the legend.
In the ergodic phase (small $h$) where the eigenstates are apparently thermal these differences vanish
as $L$ is increased, while they remain large in the localized phase (large $h$).}
\label{Sz_eq}
\end{figure}

As a simple measure to probe how \emph{thermal} the many-body eigenstates appear to be, we have looked at the expectation value of the spin:
\begin{equation}
m^{(n)}_{i\alpha} = \langle n | \hat{S}_i^z | n \rangle_{\alpha}
\end{equation}
at site $i$ in sample $\alpha$ in eigenstate $n$.  For each specific site in each sample we then compare this for eigenstates that are
adjacent in energy, showing the mean value of the difference:
$[|m^{(n)}_{i\alpha}-m^{(n+1)}_{i\alpha}|]$ for various $L$ and $h$ in Fig. 1,
where the eigenstates are labeled with $n$ in order of their energy.  The square brackets denote an average over states, samples and
sites.  The number of samples used ranges from $10^4$ for $L=8$ to 50 for $L=16$ and some values of $h$.
In our figures we show one-standard-deviation error bars.
Here and in all the data reported in this paper we restrict
our attention to the many-body eigenstates that are in the middle one-third of the
energy-ordered list of states for their sample.  Thus we are looking
only at high energy states and avoiding those states that represent low temperatures.  In this energy range, the difference in energy density
between adjacent states $n$ and $(n+1)$ is of order $h2^{-L}$ and thus exponentially small in $L$ as $L$ is increased.  If the
eigenstates are thermal then adjacent eigenstates represent temperatures that differ only by this exponentially small amount, so the expectation
value of $\hat{S}_i^z$ should be the same in these two states for $L\rightarrow\infty$.  From Fig. 1, one can see that this indeed
appears to be true in the ergodic phase at small $h$, as expected.  In the localized phase at large $h$ the differences between adjacent
eigenstates remain large as $L$ is increased, confirming that these many-body eigenstates are not thermal.

To begin to explore the possible finite-size scaling properties of the many-body localization transition in our model, we need to
measure correlations or entanglement on length scales of order the length $L$ of our samples.
One of the simplest correlation functions within a many-body eigenstate $|n\rangle$ of the Hamiltonian of sample $\alpha$ is
\begin{equation}
C_{n\alpha}^{zz}(i,j) = \langle n | \hat{S}_i^z \hat{S}_j^z | n \rangle_{\alpha} - \langle n | \hat{S}_i^z | n \rangle_{\alpha} \langle n | \hat{S}_j^z | n \rangle_{\alpha}~.
\end{equation}

\begin{figure}[!hbtp]
\includegraphics[height=2.2in,width=3.0in]{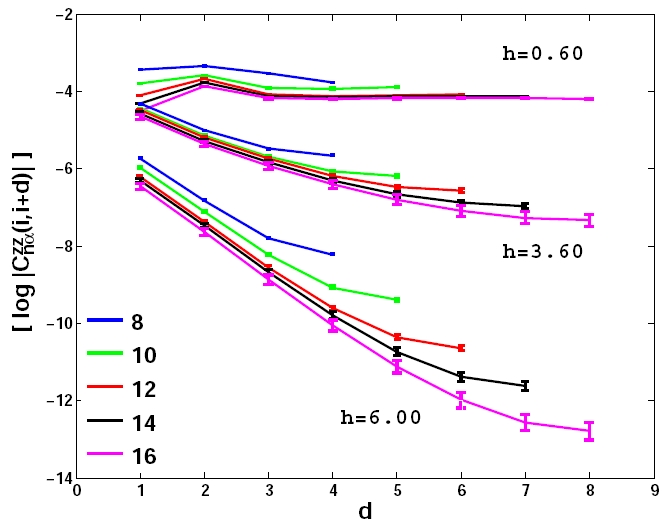}
\caption{(Color online) The spin-spin correlations in the many-body
eigenstates as a function of the distance $d$.  The sample size $L$ is
indicated in the legend.  The correlations decay exponentially with $d$ in
the localized phase ($h=6.0$), while they are independent of $d$ at large $d$
in the ergodic phase ($h=0.6$).}
\label{Czz}
\end{figure}

In Fig. 2 we show the mean value $[\log{|C_{n\alpha}^{zz}(i,i+d)|}]$ as a function of the distance $d$
for representative values of $h$ in the two phases and near
the phase transition.  Data are presented for various $L$.
This correlation function behaves very differently in the two phases:

In the ergodic phase, for large $L$ 
this correlation function
should approach its thermal equilibrium value.  For the states with zero total $\hat S^z$ that we look at, $\langle n | \hat{S}_i^z | n \rangle\cong 0$
in the thermal eigenstates of the ergodic phase.  However, the conservation of total $\hat S^z$ does result in anticorrelations so that
$C_{n\alpha}^{zz}(i,i+(L/2))\approx -1/(4(L-1))$.  These distant spins at sites $i$ and $i+(L/2)$ are entangled and correlated: if spin $i$ is flipped, that quantum
of spin is delocalized and may instead be at any of the other sites, including the most distant one.  These long-range correlations are apparent
in Fig. 2 for $h=0.6$, which is in the ergodic phase.  
Note that at large distance the correlations in the ergodic phase become essentially independent of
$d=|i-j|$ at large $L$, confirming that the spin excitations are indeed delocalized.

In the localized phase, on the other hand, the eigenstates are not thermal and $\langle n | \hat{S}_i^z | n\rangle$ remains nonzero in the limit $L\rightarrow\infty$.
If spin $i$ is flipped, within a single many-body eigenstate that quantum of spin remains localized near site $i$, with its amplitude
for being at site $j$ falling off exponentially with the distance: $C_{n\alpha}^{zz}(i,j)\sim\exp{(-|i-j|/\xi)}$, with $\xi$
the localization length.  In the localized phase the typical correlation and entanglement between two spins $i$ and $j$
thus falls off exponentially with the distance $|i-j|$ (except for $|i-j|$ near $L/2$, 
due to the periodic boundary conditions).  This behavior is apparent in Fig. 2
for $h=6.0$, which is in the many-body localized phase.  We note that in the localized phase we find that the long distance
spin correlations are of apparently random sign.


The data shown in Figs. 1 and 2 demonstrate the existence of and some of the differences between the ergodic and localized phases.
We have also looked at energy level statistics and entanglement spectra \cite{vadim} of the eigenstates (data not shown), which also support
the robust existence of these two phases. 
In addition to confirming the existence of these two distinct phases,
we would like to locate and characterize the many-body localization phase transition between them.
However, in the absence of a theory of this transition, the nature of the finite-size scaling is uncertain,
which makes it difficult to draw any strong conclusions from these data with their modest range of $L$.
In studies of ground-state quantum critical points with quenched randomness, very broadly speaking, one first step is to
classify the transitions by whether they are governed (in a renormalization group treatment) by fixed points with
finite or infinite randomness \cite{dsf,ssb}.

\begin{figure}[!hbtp]
\includegraphics[height=2.2in,width=3.0in]{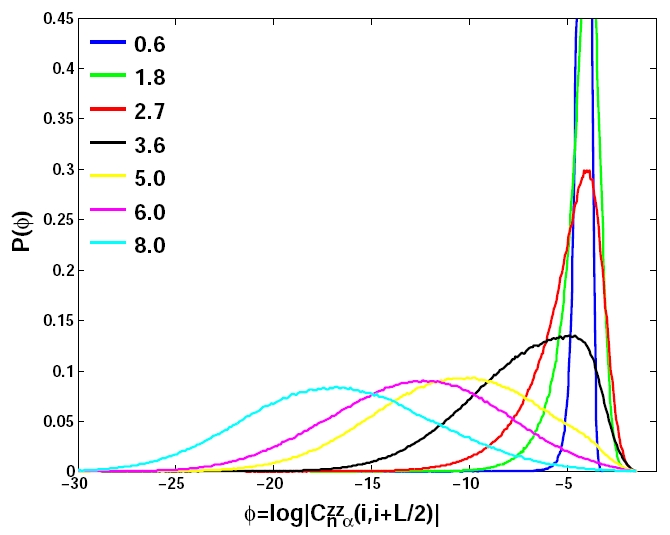}
\caption{(Color online) The probability distributions of the logarithm of the long distance
spin-spin correlation in the many-body
eigenstates for sample size $L=16$ and the values of the random field $h$
indicated in the legend.}
\label{Czz_distribution}
\end{figure}

To explore this question for our system, we look at the probability distributions of the
long distance spin correlations.  For quantum-critical ground states governed by infinite-randomness fixed points, these
probability distributions are found to be very broad \cite{dsf}.  In particular, we look at
\begin{equation}
\phi=\log{|C_{n\alpha}^{zz}(i,i+(L/2))|}~,
\end{equation}
whose probability distributions
for $L=16$ are displayed in Fig. 3 for various values of $h$.  Note the distributions are narrow, as expected, in the
ergodic phase and consistent with log-normal, as expected, in the localized phase.  In between, in the vicinity of the
apparent phase transition, the distributions are quite broad and asymmetric.

\begin{figure}[!hbtp]
\includegraphics[height=2.2in,width=3.0in]{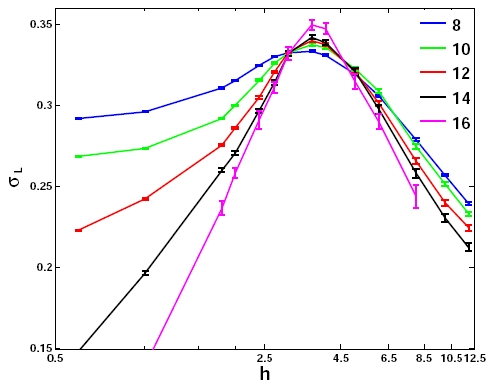}
\caption{(Color online) The scaled width $\sigma$ of the probability distribution of
the logarithm of the long-distance spin correlations (see text).
The legend indicates the sample lengths $L$.  In the ergodic phase
at small $h$ and in the localized phase at large $h$, this width decreases with increasing $L$,
while near the transition it appears to increase.  To produce the objective error-bars
shown, we have calculated the $\sigma$ (see text) for each sample by averaging only over sites and
eigenstates within each sample, and then used the sample-to-sample variations of $\sigma$
to estimate the statistical errors.  We have also (data not shown) calculated $\sigma$ by instead averaging
$\phi$ and $\phi^2$ over all samples; this produces scaling behavior for $\sigma$ that is
qualitatively the same as shown here, but with $\sigma$ somewhat larger in the localized
phase and near the phase transition.}
\label{strong_disorder}
\end{figure}

To construct a dimensionless measure of how these distributions
change shape as $L$ is increased, we divide $\phi$ by its mean, defining $\eta=\phi/[\phi]$.
Then we quantify the width of the probability distribution of $\eta$ by the standard deviation
$\sigma=\sqrt{[\eta^2]-1}$.  
This quantity is shown in Fig. 4 vs. $h$ for the various
values of $L$.  By this measure, in both the ergodic and localized phases the distributions
become narrower as $L$ is increased, as can be seen in Fig. 4.
This happens in the localized phase because although the mean of $-\phi$ grows linearly in $L$, the standard deviation is expected to
grow only $\sim\sqrt{L}$.  Over the small range of $L$ that we can explore, $\sigma$ is found to decrease more
slowly than the expected $L^{-1/2}$ in the localized phase, but it does indeed decrease.

This scaled width $\sigma_L(h)$ of the probability distribution of $\phi$ as a function of the
random field $h$ for each given sample size $L$ shows a maximum between the ergodic and localized
phases.  We take the location of this maximum as a finite-size estimate of the localization phase
transition point $h_c$, since it gets larger and sharper as $L$ is increased, suggesting that it is indeed
associated with the phase transition.  However, one should be cautious, given
that we only vary the sample size $L$ by a factor of two.  
But from Fig. 4, assuming that the maxima do estimate $h_c$, we suggest that $h_c$ is roughly 3 or 4.  
In the vicinity of the phase transition, $\sigma$ actually increases as $L$ is increased,
suggesting that its critical value is nonzero, like for quantum-critical ground states that are governed
by an infinite randomness fixed point.  This suggests the possibility that this one-dimensional many-body localization
transition might also be in an infinite-randomness universality class.

{\it Summary:}  This study of the exact many-body eigenstates of our model (1) has demonstrated some of the properties of the ergodic and localized phases.
We also find a rough estimate of the localization transition from the scaling of the probability distributions of the long-distance spin correlations. 
This scaling suggests that the 
transition might be governed by an infinite-randomness fixed point.  These results suggest that
efforts to develop a theory of this interesting phase transition should consider the
possibility of a treatment by some sort of strong disorder renormalization group approach.
Of course, the model we have studied is only one-dimensional, and the behavior of this transition in
higher dimensions might be different.

We thank Vadim Oganesyan for previous collaborations and for many useful discussions related to this work.
This work was supported by ARO Award W911NF-07-1-0464 with funds from the
DARPA OLE Program and by the NSF
through MRSEC grant DMR-0819860.


\end{document}